\documentclass[prb,twocolumn,aps,showpacs]{revtex4}
\usepackage{bm}
\usepackage{amsmath}
\usepackage{amssymb}
\usepackage[dvips]{color}
\usepackage[dvips]{graphics}
\usepackage[dvips]{graphicx}
\usepackage{eepic}
\usepackage{pstricks}%
\usepackage{pst-eps}%
\usepackage{pst-plot}%
\usepackage{pst-node}%
\usepackage{pst-poly}
\renewcommand{\d}{{\rm d}}
\renewcommand{\i}{{\rm i}}
\newcommand{\e}{{\rm e}}
\newcommand{\Tr}{{\rm Tr}\;}

\renewcommand{\Im}{{\rm Im}\;}
\newcommand{\sgn}{{\rm sgn}}

\begin{document}
\title{Quantum Hall effect in bilayer and multilayer graphenes with
finite gate voltage}
\author{Masaaki Nakamura$^1$, Lila Hirasawa$^{2,3}$,
 and Ken-Ichiro Imura$^{2}$}
\affiliation{
 $^1$Max-Planck-Institut f\"{u}r Physik komplexer Systeme,
 N\"{o}thnitzer Stra{\ss}e 38, D-01187 Dresden, Germany,\\
 $^2$Institute for Solid State Physics, University of Tokyo,
 Kashiwanoha, Kashiwa-shi, Chiba 277-8581, Japan,\\
 $^3$Department of Physics, Tokyo Institute of Technology,
 Oh-Okayama, Meguro-ku, Tokyo 152-8551, Japan}
\date{\today}
\begin{abstract}
 We discuss the quantum Hall effect of bilayer graphene with finite gate
 voltage where the Fermi energy exceeds the interlayer hopping energy.
 We calculated magnetic susceptibility, diagonal and off-diagonal
 conductivities in finite-magnetic-field formalism, and observed
 crossover of integer quantum Hall effect from two independent monolayer
 type system to strongly coupled bilayer systems by changing the ratio
 of interlayer hopping energy and the gate voltage.  We also discuss the
 case of multilayer systems with Bernal stacking.
\end{abstract}

\maketitle

\smallskip

\begin{figure}[h!]
 \begin{center}
  \psset{unit=7mm}
  \begin{pspicture}(0,-2.0)(18,2.5)
   \rput(3,0){
   \rput[l](-2.8,-2.0){(a)monolayer}
   \rput[r](-0.2,2.4){$\sigma_{xy}$}
   \rput[r](2.5,-0.4){$1/B$}
   \psaxes[labels=none,ticks=none]{->}(0,0)(-2.5,-2.5)(2.5,2.5)
   \psline[linecolor=red]%
   (-2,1.75)(-1.5,1.75)(-1.5,1.25)(-1.0,1.25)(-1.0,0.75)
   (-0.5,0.75)(-0.5,0.25)(0,0.25)(0.0,-0.25)(0.5,-0.25)(0.5,-0.75)
   (1.0,-0.75)(1.0,-1.25)(1.5,-1.25)(1.5,-1.75)(2.0,-1.75)
   }
   \rput(9,0){
   \rput[l](-2.8,-2.0){(b)bilayer}
   \rput[r](-0.2,2.4){$\sigma_{xy}$}
   \rput[r](2.5,-0.4){$1/B$}
   \psaxes[labels=none,ticks=none]{->}(0,0)(-2.5,-2.5)(2.5,2.5)
   \psline[linecolor=red]%
   (-2,2)(-1.5,2)(-1.5,1.5)(-1.0,1.5)(-1.0,1.0)(-0.5,1.0)(-0.5,0.5)(0,0.5)
   (0.0,-0.5)(0.5,-0.5)(0.5,-1.0)(1.0,-1.0)(1.0,-1.5)(1.5,-1.5)(1.5,-2)(2.0,-2)
   \psline{<->}(-1.3,1.0)(-1.3,0.5)
   \psline{-}(-1.5,1.0)(-1.1,1.0)
   \psline{-}(-1.5,0.5)(-1.1,0.5)
   \rput[r](-1.7,0.75){$e^2/h$}
   }
%
%
   \psset{unit=3mm,linewidth=0.3pt}
   \rput(10,2.7){
     \psline(-1.5,+2.0)(0,0)(1.5,+2.0)
     \pscustom{
     \psline(-1.5,-2.0)(0,0)(1.5,-2.0)
     \gsave
     \fill[fillstyle=solid,fillcolor=gray]
     \grestore
     }
     \pscustom{
     \psline(-0.3,0.4)(0,0)(0.3,0.4)
     \gsave
     \fill[fillstyle=solid,fillcolor=gray]
     \grestore
     }
     \rput[r](-0.2,2.4){$E$}
     \rput[r](2.5,-0.5){$k$}
     \psaxes[labels=none,ticks=none]{->}(0,0)(-2.5,-2.5)(2.5,2.5)
     }
     \rput(24,2.7){
     \pscustom{
     \psplot[plotstyle=curve]{-0.55}{0.55}{x 2 exp}
     \gsave
     \fill[fillstyle=solid,fillcolor=gray]
     \grestore
     }
     \pscustom{
     \psplot[plotstyle=curve]{-1.414}{1.414}{x 2 exp -1 mul}
     \gsave
     \fill[fillstyle=solid,fillcolor=gray]
     \grestore
     }
     \psplot[plotstyle=curve]{-1.414}{1.414}{x 2 exp}
     \rput[r](-0.2,2.4){$E$}
     \rput[r](2.5,-0.5){$k$}
     \psaxes[labels=none,ticks=none]{->}(0,0)(-2.5,-2.5)(2.5,2.5)
     }
  \end{pspicture}
 \end{center}
 \caption{(Color online) Schematic illustration of the anomalous integer
 quantum Hall effect (per valley and per spin) of (a) monolayer and (b)
 bilayer graphenes. The twice larger step at $1/B=0$ in (b) is due to
 the two-fold degeneracy of the Landau level at zero energy. The insets
 show dispersion relations of both systems.}\label{illustration_of_QHE}
\end{figure}
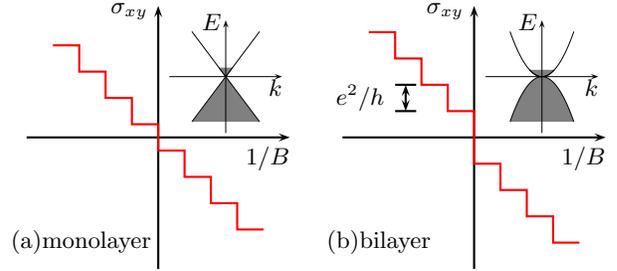

{\em Introduction}---
Among numbers of unusual physical properties of graphene (atomically
thin graphite),\cite{Novoselov,Zhang} study of the anomalous integer
quantum Hall effect (QHE) was crowned with dramatic success.  The Hall
conductivity (per valley and per spin) is quantized as
$\sigma_{xy}=-\frac{e^2}{h}(n+1/2)$, $n=0,\pm 1,\pm 2,\cdots$ wheres
$\sigma_{xy}=-\frac{e^2}{h}n$ for usual two-dimensional (2D) electron
gas.  The difference $1/2$ is explained theoretically, based on the 2D
massless Dirac fermions where a Landau level is located at zero energy.
\cite{Zheng-A,Gusynin-S_2005b,Gusynin-S_2006,Peres-G-C} Moreover, QHE in
bilayer system has also been observed as $\sigma_{xy}=-\frac{e^2}{h} n$,
$n=\pm 1,\pm 2,\cdots$.\cite{Novoselov} The characteristic feature of
this result is that the step of $\sigma_{xy}$ at the
strong-magnetic-field limit is twice larger than other steps (see
Fig.~\ref{illustration_of_QHE}). This behavior was successfully
explained by McCann and Fal'ko based on an effective Hamiltonian in
$2\times 2$ matrix form.\cite{McCann-F}

On the other hand, one of the special situations of graphene system
which can not be realized in other Dirac fermion systems such as
organic conductor $\alpha$-(BEDT-TTF)$_2$I$_3$\cite{Katayama-K-S} is
that the Fermi energy is tunable parameter by the gate voltage. Although
the theory by McCann and Fal'ko well describes QHE in sufficiently small
Fermi-energy regions, their Hamiltonian is no longer valid when the gate
voltage becomes greater than the energy gap between two bands, which is
the same order of the interlayer hopping energy.  This situation is
recently realized experimentally.\cite{Henriksen} For such cases, we
need to discuss the QHE based on the Hamiltonian of the bilayer graphene
in $4\times 4$ matrix form [eq.~(\ref{Ham_bilayer.3}) below].

The bilayer Hamiltonian is also important to discuss multilayer systems.
For the Bernal stacking structure (it is also called AB or staggered
stacking, and about 80 \% of natural graphite falls into this category),
the Hamiltonian of an $N$-layer system given by $2N\times 2N$ matrix can
be block diagonalized into effective bilayer systems and a monolayer if
$N$ is odd.\cite{Guinea-C-P,Koshino-A_2007b,Nakamura-H,Min-M} This
decomposition is related to the Fourier modes of the wave function along
the stacking direction.\cite{Koshino-A_2007b} In this paper, we discuss
QHE of bilayer system with finite gate voltage based on the four band
Hamiltonian (\ref{Ham_bilayer.3}), and also that of multilayer systems.

\smallskip

{\em Eigenvalues and eigenstates}---
The Hamiltonian of the bilayer graphene in $4
\times 4$ matrix form is given by
\begin{equation}
 \mathcal{H}
  =\left[
    \begin{array}{cccc}
     0      & v\pi_- & 0      & t      \\
     v\pi_+ &      0 & 0      & 0      \\
     0      &      0 & 0      & v\pi_- \\
     t      &      0 & v\pi_+ & 0
    \end{array}
\right],
  \label{Ham_bilayer.3}
\end{equation}
where $\pi_{\pm}\equiv\pi_x\pm\i\pi_y$ with
$\bm{\pi}\equiv\bm{p}+e\bm{A}/c$ is the momentum operator in a magnetic
field $\nabla\times\bm{A}=(0,0,B)$.  $v$ and $t$ are the Fermi velocity
and interlayer hopping energy, respectively. We have ignored the
trigonal wrapping effect stems from the next-nearest-interlayer hopping
for simplicity.  For the system with zero-magnetic field ${\cal H}_0$,
dispersion relation is $E_{\mu}(\bm{k})=s_2(\sqrt{t^2+(2v\hbar k)^2}+s_1
t)/2$, where the label $\mu=(s_1,s_2)$ specifies the outer and the inner
bands ($s_1=\pm 1$), and positive and negative ($s_2=\pm 1$) energies,
respectively, so that $t$ corresponds to an energy gap between two
bands.  Since the commutation relation between the momentum operators in
eq.~(\ref{Ham_bilayer.3}) is $[\pi_{\pm},\pi_{\mp}]=\mp 2eB\hbar/c$,
there are correspondences with the creation and annihilation operators
of the harmonic oscillator:
$\pi_{\pm}\to\sqrt{2}\frac{\hbar}{l}a^{\dag}$ $\pi_{\mp}\to
\sqrt{2}\frac{\hbar}{l}a$ for $eB\gtrless 0$, where
$l\equiv\sqrt{c\hbar/|eB|}$. Assuming that the wave function is given by
linear combination of the number states of the harmonic oscillator
$|n\rangle$, we obtain eigenvalues and eigenstates of
eq.~(\ref{Ham_bilayer.3}) as follows: The eigenvalues are
\begin{subequations}
\begin{align}
 &
 E_n^\mu=\frac{\sqrt{2}\hbar v}{l}\lambda_n^\mu,\qquad \mu=(s_1,s_2),\\
 &
 \lambda_n^\mu=
  s_2\sqrt{\frac{2n+1+r^2+s_1\sqrt{r^4+2(2n+1)r^2+1}}{2}},
  \label{eigenvalues}
\end{align}
\end{subequations}
where $r\equiv\frac{l}{\sqrt{2}\hbar v}t$. 
For $n\geq 0$ with $\lambda\neq 0$, the eigenstates are given by
\begin{subequations}
\begin{equation}
 |n,\mu\rangle\rangle=
  \left[
   \begin{array}{c}
    \alpha_n^{\mu}|n\rangle\\
    \beta_{n+1}^{\mu}|n+1\rangle\\
    \gamma_{n-1}^{\mu}|n-1\rangle\\
    \delta_n^{\mu}|n\rangle
   \end{array}
  \right],\
  \left[
   \begin{array}{c}
    \alpha_n^{\mu}\\
    \beta_{n+1}^{\mu}\\
    \gamma_{n-1}^{\mu}\\
    \delta_n^{\mu}
   \end{array}
  \right]
  =\alpha_n^{\mu}
  \left[
   \begin{array}{c}
    1\\
    \frac{\sqrt{n+1}}{\lambda_{n,\mu}}\\
    \frac{\sqrt{n}r}{\lambda_{n,\mu}^2-n}\\
    \frac{\lambda_{n,\mu}r}{\lambda_{n,\mu}^2-n}
   \end{array}
  \right],
\end{equation}
\begin{equation}
 \alpha_n^{\mu}
   =\left[1+\frac{n+1}{\lambda_{n,\mu}^2}
     +\frac{(\lambda_{n,\mu}^2+n)r^2}{(\lambda_{n,\mu}^2-n)^2}
    \right]^{-1/2}.
\end{equation}
\end{subequations}
The zero-energy state ($\lambda=0$) is doubly degenerate.  For these
states, we specify the quantum numbers $n=0$ and $\mu$ as follows,
\begin{equation}
 |0,-,+\rangle\rangle\equiv
  \frac{1}{\sqrt{1+r^2}}
  \left[
   \begin{array}{l}
    0\\
    r|1\rangle\\
    0\\
    -1|0\rangle
   \end{array}
  \right],\
 |0,-,-\rangle\rangle\equiv
  \left[
   \begin{array}{l}
    0\\
    |0\rangle\\
    0\\
    0
   \end{array}
  \right].
\end{equation}

For large interlayer hopping $r^2\gg 1$, using the Taylor expansion, the
inner band $s_1=-1$ is approximated as
\begin{equation}
  E_n^{-,\pm}=\pm\frac{\sqrt{2}}{t}\frac{\hbar^2v^2}{l^2}\sqrt{n(n+1)}.
  \label{ene_low}
\end{equation}
This is consistent with the result obtained by McCann and
Fal'ko.\cite{McCann-F} Since the obtained eigenstates are given by
number states like the nonrelativistic free fermion system, the
degeneracy of each Landau level is also discussed in the same way that
this is given by multiplicity of center of the coordinate: $V/2\pi l^2$
with $V$ being volume of the system.

\smallskip

{\em Susceptibility}---
Before discussing the transport properties, we derive the magnetic
susceptibility based on the the finite-magnetic-field formalism to check
the consistency with other formalisms. In fact, the Hamiltonian
(\ref{Ham_bilayer.3}) was introduced two decades ago by Safran motivated
by graphite intercalation compounds.\cite{Safran} He calculated magnetic
susceptibility based on the weak-magnetic field formalism. An extention
of this result with impurity scattering $\Gamma=\hbar/2\tau$ with $\tau$
being the collision time of quasiparticles is given by\cite{Nakamura-H}
\begin{subequations}
\begin{align}
 &
 \chi=
 \frac{e^2 v^2}{12\pi^2 c^2}
 \int_{-\infty}^{\infty}\d x f(x)
 \Im F(x+\i\Gamma),
 \label{sus_bilayer.1}\\
 &
 F(x)=
 -\frac{3}{tx}\log\frac{x+t}{x-t}
 +\frac{2}{x^2-t^2},
 \label{sus_bilayer.2}
\end{align}
\end{subequations}
where $f(x)\equiv(\e^{\beta(x-\mu)}+1)^{-1}$ is the Fermi distribution
function with $\beta\equiv 1/k_{\rm B}T$ being inverse temperature.  As
discussed by Safran, in $\Gamma\to 0$ limit, the first term of the r.h.s
of eq.~({\ref{sus_bilayer.2}) gives diamagnetic logarithmic divergence
near the zero Fermi energy $|\mu|\to 0$, while the second term gives
paramagnetic behavior around $|\mu|\sim t$.  Now let us calculate the
magnetic susceptibility by the Landau quantization formalism.  According
to the functional integral method, the thermodynamic potential is given
by
\begin{align}
\lefteqn{
 \Omega(B)
 =-\frac{1}{\beta}\sum_{n=-\infty}^{\infty}
 \Tr\ln(-\i\tilde{\omega}_n+\hat{\cal H}_0/\hbar)}\label{thermo.potential}\\
 =&-\frac{V}{2\pi l^2\beta}
 \sum_{n=-\infty}^{\infty}
 \sum_{s_1=\pm}
 \sum_{k=0}^{\infty}
 \ln\left[
 (\i\tilde{\omega}_n)^2-(E_{k}^{s_1,\pm}/\hbar)^2
 \right].\nonumber
\end{align}
Here $\tilde{\omega}_{n}$ is Matsubara frequency of fermion including
the chemical potential $\mu$ and effect of impurity scattering $\Gamma$
as $\i\tilde{\omega}_{n}=\i\omega_n + [\mu + \i\
\sgn(\omega_n)\Gamma]/\hbar$.  Applying the Euler-Maclaurin formula
\begin{equation}
 \sum_{k=a}^{b-1}g\left(k+\frac{1}{2}\right)\simeq
  \int_a^b g(x)\d x-\frac{1}{24}[g'(b)-g'(a)],
  \label{Euler-Maclaurin}
\end{equation}
to the second derivative of eq.~(\ref{thermo.potential}): $\chi=
-\frac{1}{V}\left.\frac{\partial^2 \Omega}{\partial B^2}\right|_{B=0}$,
the first (second) term of the r.h.s of eq.~(\ref{Euler-Maclaurin})
gives the first (second) term of the r.h.s of
eq.~({\ref{sus_bilayer.2}).  Thus the result of this calculation
coincides with that of the weak-magnetic field formalism.

\smallskip

{\em Conductivity}---
The conductivity is given by the Kubo formula as,
\begin{equation}
 \sigma_{ij}(\Omega)=
  \frac{\Im\tilde{\Pi}_{ij}(\Omega+\i\eta)}{\hbar\Omega},
\end{equation}
where $\tilde{\Pi}_{ij}(\Omega)\equiv \Pi_{ij}(\Omega)-\Pi_{ij}(0)$ with
$\{i,j\}\in\{x,y\}$.  The polarization function $\Pi_{ij}(\Omega)$ is
given by the current-current correlation function, and obtained as the
analytical continuation of the Matsubara form:
\begin{align}
\lefteqn{\tilde{\Pi}_{ij}(\i\nu_m)=}
 \label{hall_conductivity_xx_derivation2}\\
&-\frac{e^2}{2\pi l^2\beta\hbar}\sum_{n=-\infty}^{\infty}
 \sum_{k,l}\sum_{\mu,\nu}
 \frac{
 \langle\langle k,\mu|\gamma_i|l,\nu\rangle\rangle
 \langle\langle l,\nu|\gamma_j|k,\mu\rangle\rangle
 }{[\i\tilde{\omega}_{n}-\tilde{E}_k^{\mu}]
 [\i\tilde{\omega}_{n}+\i\nu_m-\tilde{E}_l^{\nu}]},
\nonumber
\end{align}
where the matrix $\gamma_{i}$ is defined by $\bm{\gamma}\equiv
\nabla_{\bm{k}}{\cal H}_0/\hbar$, and $\tilde{E}_k^{\mu}\equiv
E_k^{\mu}/\hbar$. $\i\nu_m$ is Matsubara frequency of boson.  We have
ignored vertex corrections.  After calculating the Matsubara frequency
sums, and matrix elements $\langle\langle\cdots\rangle\rangle
\langle\langle\cdots\rangle\rangle$ based on their symmetric
(antisymmetric) properties for $\sigma_{xx}$ ($\sigma_{xy}$), the
general expression of the conductivity (per valley and per spin) is
obtained in the unified form as
\begin{align}
\lefteqn{\sgn(eB)\sigma_{xy}(\Omega)+\i\sigma_{xx}(\Omega)}
 \label{conductivity.general}
\\
 &=-\frac{e^2}{h}\frac{2v^2}{l^2}
 \biggl[
 \sum_{k\geq 1}\sum_{\mu,\nu}
 X_{k+1,k}^{\mu,\nu}
 (\alpha_{k+1}^{\mu}\beta_{k+1}^{\nu}+\gamma_{k}^{\mu}\delta_{k}^{\nu})^2
 \nonumber\\
 &
 +\sum_{\mu,s_2=\pm}
 X_{1,0}^{\mu,(+,s_2)}
 (\alpha_{1}^{\mu}\beta_{1}^{+,s_2}
 +\gamma_{0}^{\mu}\delta_{0}^{+,s_2})^2\nonumber\\
 &
 +\sum_{\mu}
 X_{1,0}^{\mu,(-,+)}
 \frac{(r\alpha_1^{\mu}-\gamma_0^{\mu})^2}{1+r^2}
 +\sum_{s_2=\pm}
 X_{0,0}^{(+,s_2),(-,-)}(\alpha_{0}^{+,s_2})^2\biggr].
\nonumber
\end{align}
Here the matrix elements between Landau levels $k$ and $k+1$ remain for
$k\geq 1$, but other elements involving $k=0$ states are complicated due
to the degeneracy of zero energy state.  We have defined
\begin{align}
 X_{k,l}^{\mu,\nu}(\Omega)
 \equiv
 &
 \sum_{n}\Bigl[
 \left.
 (\i\tilde{\omega}_n-\tilde{E}_k^{\mu})^{-1}
 (\i\tilde{\omega}_n+\i\nu_m-\tilde{E}_l^{\nu})^{-1}
 \right|_{\i\nu_m\to\Omega}
 \nonumber\\
 &
 -(\Omega\to-\Omega)
 \Bigr](\Omega\beta\hbar)^{-1},
\end{align}
and evaluated it analytically.

\smallskip

{\em Quantum Hall effect}---
For clean system $\Gamma=0$ and dc limit $\Omega\to 0$,
$X_{k,l}^{\mu,\nu}=[f(E_k^{\mu})-f(E_l^{\nu})]/
(\tilde{E}_k^{\mu}-\tilde{E}_l^{\nu})^2$.  Then after straightforward
calculations of eq.~(\ref{conductivity.general}), expression of the Hall
conductivity becomes a simple form,
\begin{align}
\lefteqn{\sigma_{xy}=-\sgn(eB)\frac{e^2}{h}
 \biggl[
 \sum_{k\geq 1}\sum_{\mu}
 \tilde{f}(E^{\mu}_{k})
 +\sum_{s_2=\pm} \tilde{f}(E^{+,s_2}_{0})}\nonumber\\
&
 +\left(2-\frac{1}{1+r^2}\right)\tilde{f}(E^{-,+}_0)
 +\frac{1}{1+r^2}\tilde{f}(E^{-,-}_0)
 \biggr].
 \label{sxy.4}
\end{align}
Here, we have defined $\tilde{f}(x)\equiv f(x)-1/2$. In this derivation
it is important to keep the formula to satisfy the particle-hole
symmetry.\cite{Gusynin-S_2005b,Gusynin-S_2006} Moreover, assuming zero
temperature and positive Fermi energy $\mu>0$, we have
\begin{align}
 \lefteqn{\sigma_{xy}=
 -\sgn(eB)\frac{e^2}{h}
 \sum_{k\geq 0}\sum_{s_1}\theta(\mu-E_k^{s_1,+})}\\
 =&-\sgn(eB)\frac{e^2}{h}
 \biggl\{
 \biggl[
 \frac{\mu^2+\sqrt{\mu^2 t^2+\tilde{B}^2}}{2|\tilde{B}|}
 +\frac{1}{2}\biggr]_G\label{sxy.6}\\
 &
 +
 \theta(\mu^2-t^2-2|\tilde{B}|)
 \biggl[
 \frac{\mu^2-\sqrt{\mu^2 t^2+\tilde{B}^2}}{2|\tilde{B}|}
 +\frac{1}{2}
 \biggr]_G
 \biggr\}\nonumber
\end{align}
where $[x]_G$ means the integer part of $x$, and we have defined
$\tilde{B}\equiv\hbar v^2eB/c$. The first and the second terms of
eq.~(\ref{sxy.6}) correspond to contribution from the inner ($s_1=-1$)
and that from the outer ($s_1=+1$) bands, respectively.  As expected,
the Hall conductivity is quantized as a unit of $e^2/h$, reflecting that
it is a topological number. Although, for the real QHE, the Anderson
localization due to the impurity scattering is essential, the present
calculation with $\Gamma=0$ is useful to estimate possible step
structures of $\sigma_{xy}$.

\begin{figure}
 \includegraphics[height=4.5cm]{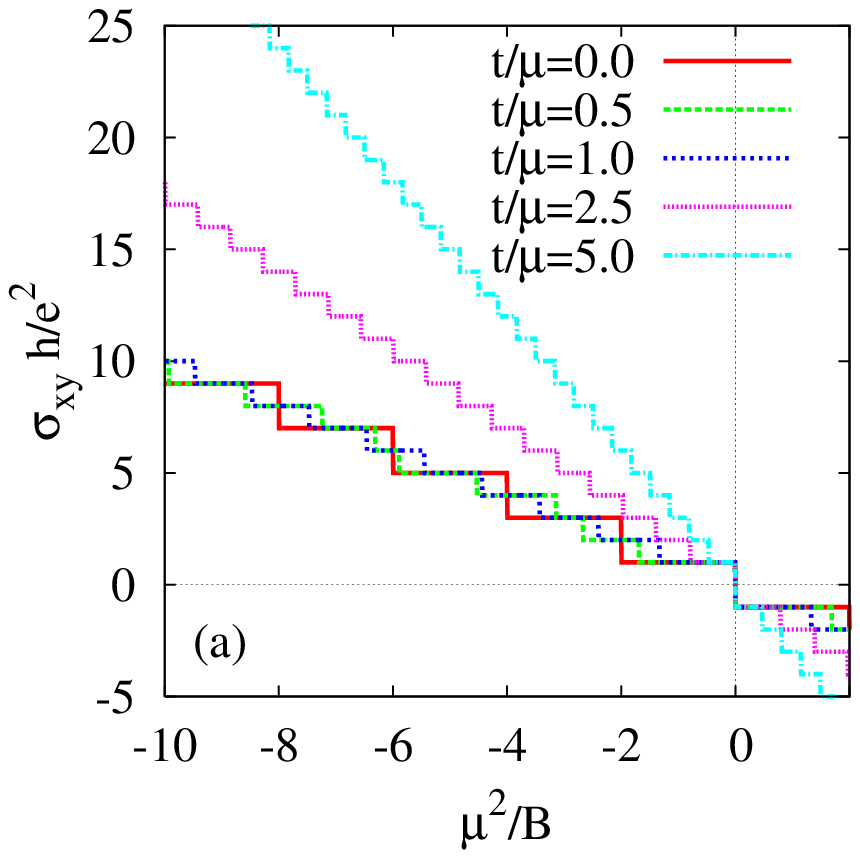}
 \includegraphics[height=4.5cm]{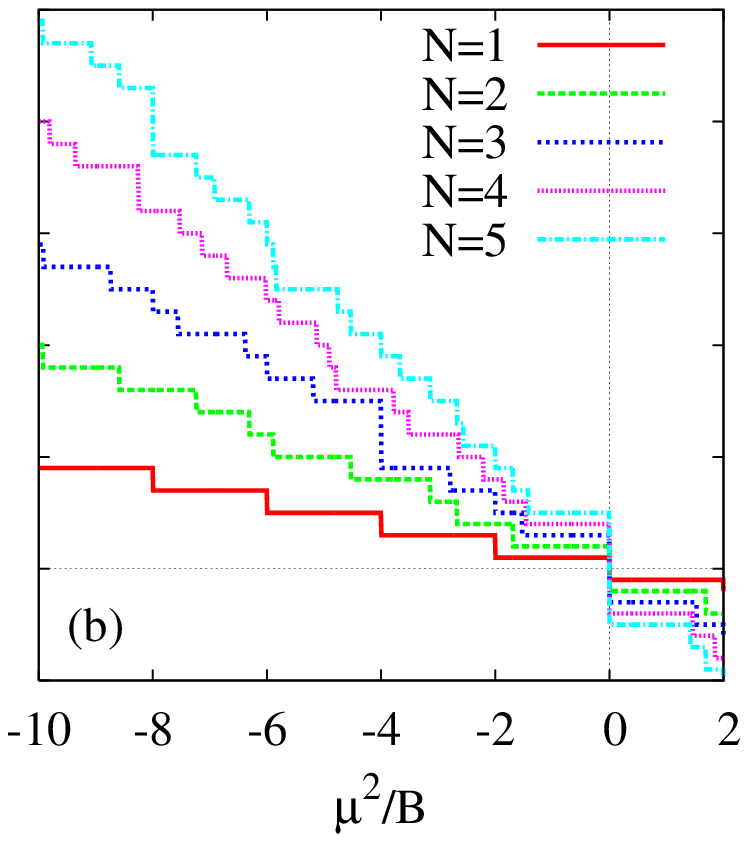}
 \caption{(Color online) Quantized Hall conductivity for $\Gamma=0$
 [eq.~(\ref{sxy.6})] versus inverse magnetic field $1/B$ for (a) bilayer
 systems and (b) $N$-layer systems ($N=1$-$5$) with Bernal stacking and
 $t/\mu=0.5$.}  \label{fig:sigma_xy_steps}
\end{figure}

\begin{figure*}
\begin{center}
 \includegraphics[height=4cm]{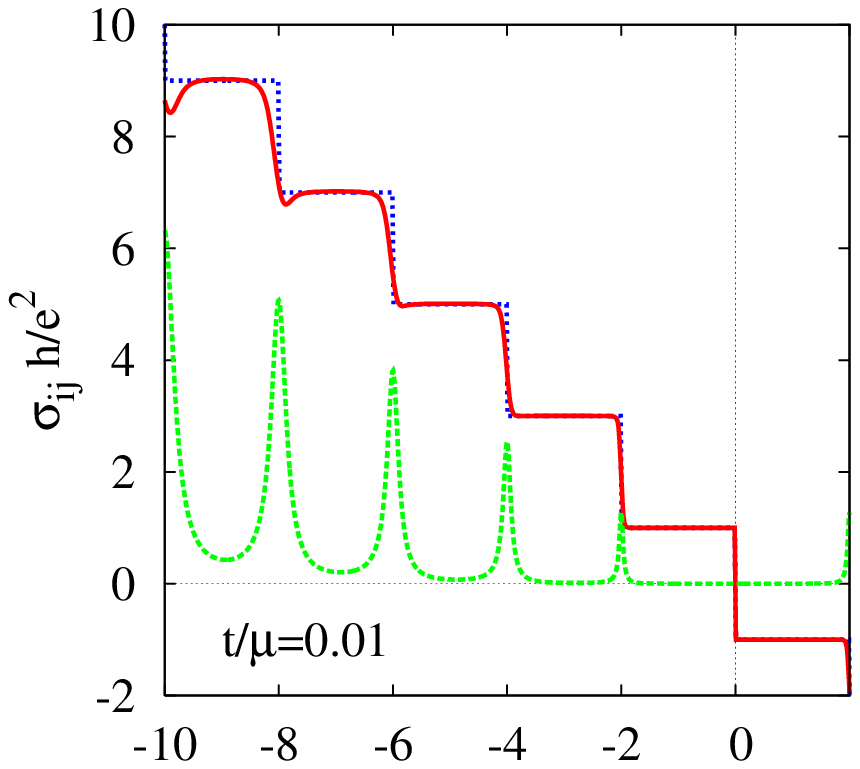}
 \includegraphics[height=4cm]{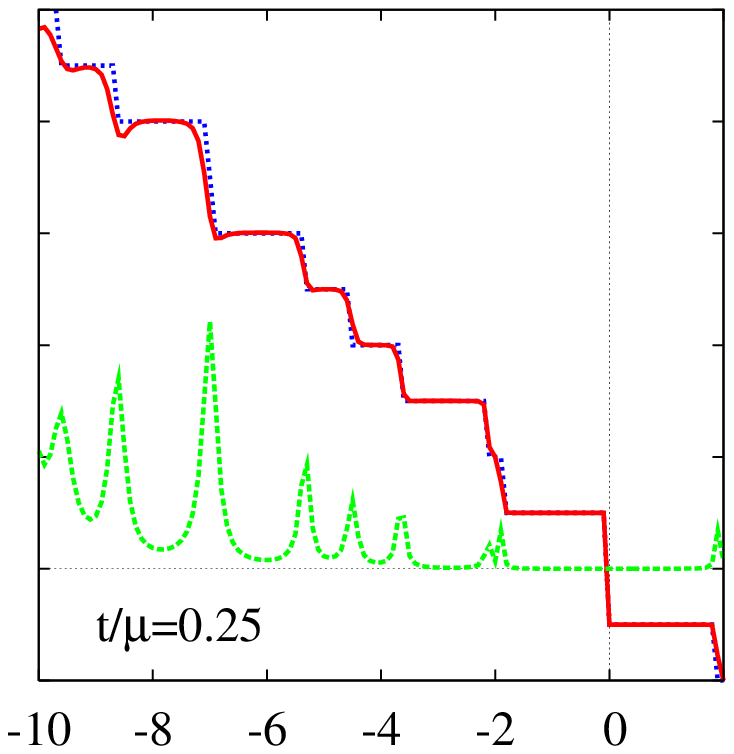}
 \includegraphics[height=4cm]{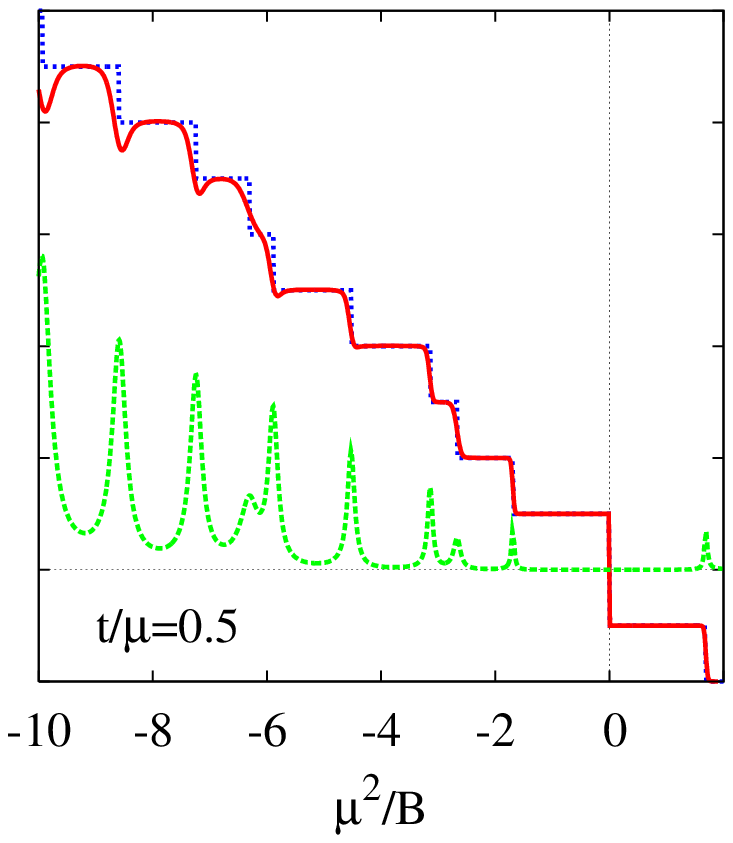}
 \includegraphics[height=4cm]{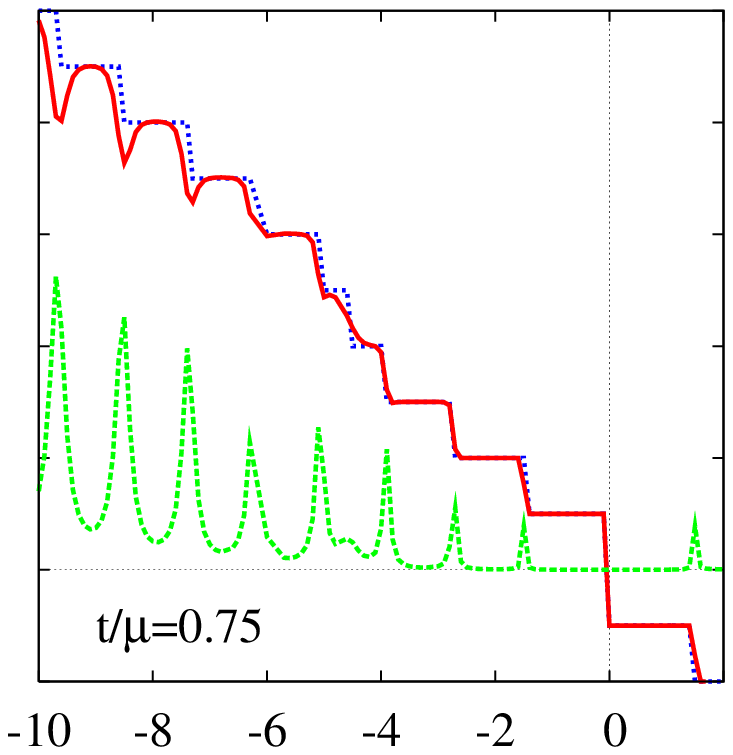}
 \includegraphics[height=4cm]{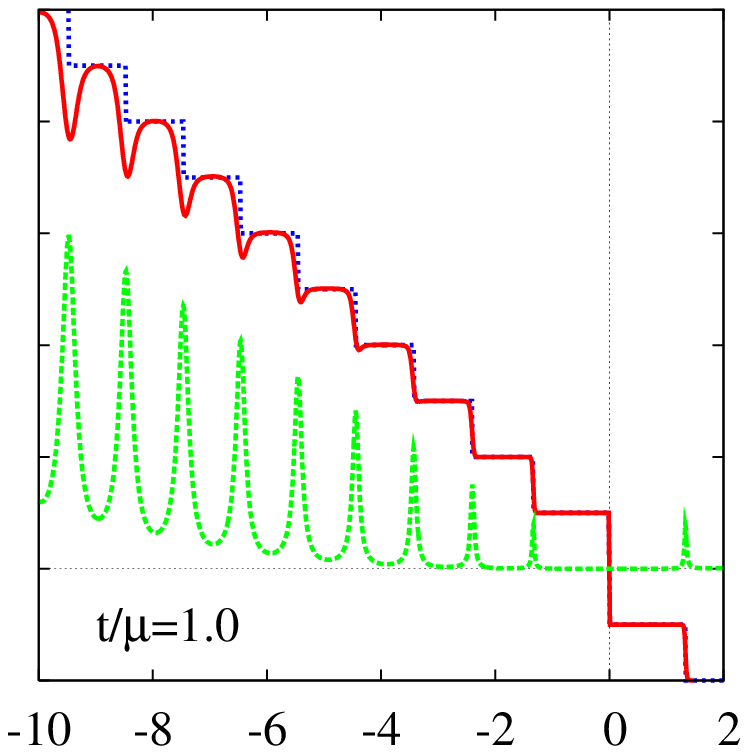}
\end{center}
 \caption{(Color online) Crossover of quantum Hall effect from two
 independent monolayer like regime to strongly coupled bilayer regime,
 observed as changing the ratio of interlayer hopping energy $t$ and the
 gate voltage (Fermi energy $\mu$).  The red and green lines denote
 off-diagonal ($\sigma_{xy}$) and diagonal ($\sigma_{xx}$)
 conductivities with $\Gamma/\mu=0.01$, respectively. The blue lines
 indicate $\sigma_{xy}$ with $\Gamma=0$.} \label{fig:sigma_xy_mu}
\end{figure*}

\smallskip

{\em Results}---
The Hall conductivity $\sigma_{xy}$ with $\Gamma=0$ versus inverse
magnetic field $1/B$ at zero temperature [eq.~(\ref{sxy.6})] is shown in
Fig.~\ref{fig:sigma_xy_steps}(a). Although the tunable parameter by gate
voltage is $\mu$ rather than $t$ in experiment, we have shown the data
as fixed $t$ for later convenience. The results show clear crossover
behavior from two independent monolayer like regime to strongly coupled
bilayer regime as the ratio of $t/\mu$ is changed.  Similar structures
of $\sigma_{xy}$ with inhomogeneous steps is also suggested to appear in
monolayer system with interaction between different $K$ points.
\cite{Ostrovsky-G-M} In these data, we find that the slope of the Hall
conductivity with respect to $\mu^2/B$ changes for $|t/\mu|>1$. This
phenomenon is explained by rewriting inside of $[\cdots]_G$ in
eq.~(\ref{sxy.6}) as
\begin{equation}
 \cdots=\frac{\mu^2}{2|\tilde{B}|}
  \biggl(
  1-s_1\sqrt{\frac{t^2}{\mu^2}+\frac{\tilde{B}^2}{\mu^4}}
  \biggr)
  +\frac{1}{2}.
   \label{slope_u}
\end{equation}
In r.h.s of eq.~(\ref{slope_u}), $(\cdots)$ is $1+O(|t/\mu|)$ for
$|t/\mu|<1$, which means that the slope of conductivity is proportional
to $\mu^2/\tilde{B}$. On the other hand, for $t/\mu>1$, the energy band
related to the QHE is only the inner band ($s_1=-1$), and $(\cdots)$
becomes larger than $1$ which means that the Landau levels strongly
depend on the interlayer hopping $t$.

Next, in Fig.~\ref{fig:sigma_xy_mu}, we show $1/B$ dependence of the
diagonal $\sigma_{xx}$ and the off-diagonal $\sigma_{xy}$ conductivities
with impurity scattering $\Gamma=0.01\mu$.  We have calculated dc limit
of eq.~(\ref{conductivity.general}) analytically, except for summation
over the Landau levels. In principle, this summation can also be
performed analytically using the Poisson formula.\cite{Shirasaki-E-H-N}
We find dip structure in the steps in the region $|\sigma_{xy}|\gtrsim
5e^2/h$.  The reason for this phenomenon is considered that the
expression of $\sigma_{xy}$ implies that of $\sigma_{xx}$, and
contribution of $\sigma_{xx}$ becomes large as the magnetic field is
decreased. Actually, in the classical result for nonrelativistic free
fermions, the expression is $\sigma_{xy}=-nec+\frac{1}{\omega_{\rm
c}\tau}\sigma_{xx}$ where $n$ and $\omega_{\rm c}\propto eB$ are the
electron density and the cyclotron frequency, respectively. This
relation is also obtained in the analytical expression of the quantum
case of 2D free fermions with constant $\Gamma$.\cite{Shirasaki-E-H-N}

Finally, we consider multilayer systems.  Since Hamiltonians of
multilayer systems with Bernal stacking are known to be decomposed into
$[N/2]_G$ bilayer systems with effective hopping
$t^*=t\sin\frac{m\pi}{2(N+1)}$ [$m=-(N-1),-(N-3),\cdots,N-1$], and a
monolayer if $N$ is
odd,\cite{Guinea-C-P,Koshino-A_2007b,Nakamura-H,Min-M} the Hall
conductivity can be calculated combining the above results.  For
example, the Hall conductivity with $\Gamma=0$ of $N$-layer system
($N=1$-$5$) obtained by eq.~(\ref{sxy.6}) and this matrix decomposition
technique with $t/\mu=0.5$ is shown in
Fig.~\ref{fig:sigma_xy_steps}(b). Although QHE of multilayer systems
with general closed packed stacking structure where Bernal and
rhombohedral stacking are mixed is discussed by Min and MacDonald using
the matrix decomposition technique and the $2\times2$ effective
Hamiltonians,\cite{Min-M} their results are limited to the small gate
voltage regions where only the inner band contributes to $\sigma_{xy}$.

\smallskip

{\em Summary}---
We have discussed physical properties of bilayer graphene in finite
magnetic field and finite gate voltage based on the Hamiltonian with
four energy bands.  We have checked the consistency between the weak-
and the finite-magnetic field formalisms calculating the magnetic
susceptibility. Then general formula of conductivity in this systems has
been obtained. We have observed crossover of integer quantum Hall effect
from two independent monolayer system to strongly coupled bilayer
systems by changing the ratio of interlayer hopping energy and the gate
voltage.

\smallskip

{\em Acknowledgments}---
The authors are grateful to M.~Oshikawa for discussion.  M.~N. thanks
R.~Shirasaki, A.~Endo, N.~Hatano, and H.~Nakamura for discussion about
their works.\cite{Shirasaki-E-H-N}

\smallskip

After this paper has almost been completed, we became aware of the paper
by M.~Koshino and T.~Ando, Phys. Rev. B {\bf 77}, 115313 (2008), which
uses the same $4\times 4$ Hamiltonian as ours to discuss the optical
properties at zero gate voltage.



\end{document}